\documentclass[prd,aps,nofootinbib,superscriptaddress]{revtex4}
\usepackage{epsfig}
\usepackage{amsmath, slashed}
\usepackage{xcolor}
\usepackage{appendix}
\usepackage{dsfont}

\usepackage[normalem]{ulem}

\usepackage{bm}
\usepackage{amsfonts}

\allowdisplaybreaks
\begin{document}

\title{Kinematical  higher-twist
corrections in $\gamma^*  \to M_1  M_2 \gamma$: \\  Charged meson production}

\author{Bernard Pire }
\email[]{bernard.pire@polytechnique.edu}
\affiliation{CPHT, CNRS, \'Ecole polytechnique, Institut Polytechnique de Paris, 91128 Palaiseau, France}
\author{Qin-Tao Song}
\email[Corresponding author: ]{songqintao@zzu.edu.cn}
\affiliation{School of Physics and Microelectronics, Zhengzhou University, Zhengzhou, Henan 450001, China}

\date{\today}

\begin{abstract}
{Generalized distribution amplitudes (GDAs) of mesons can be probed by the reactions $e^- e^+   \to M_1 M_2 \gamma$, which are accessible at electron-positron colliders such as BESIII and Belle II. After discussing the neutral meson production case in the first paper of this series \cite{Pire:2023kng}, we discuss here the complementary case of  the charged meson ($M^+ M^-$)  production, where one can extract the complete information on GDAs from  the interference of the amplitudes of the two competing processes where the photon is emitted either in the initial or in the final state.
Considering the importance of the charged meson production, we present a complete expression for the interference term of the cross section which is experimentally accessible thanks to its charge conjugation specific property.
We  adopt two types of models for  leading twist $\pi \pi$ GDAs to estimate
the size of the interference term in the process $e^- e^+   \to \pi^+ \pi^- \gamma$
numerically, namely a model  extracted from previous experimental results on $\gamma^* \gamma \to \pi^0 \pi^0$ at Belle and the  asymptotic  form predicted by QCD evolution equations.
We include in the calculation the   kinematical power suppressed (sometimes called kinematical higher-twist)  corrections  up to   $1/Q^2$  for the helicity amplitudes.
Both models of GDAs indicate that the kinematical corrections are not negligible for the interference term of the cross section measured at BESIII, thus it is necessary to  include them if we try to extract the GDAs precisely. On the other side, the  kinematical corrections are very tiny for the measurements at Belle II, and the leading twist-2 formula of the interference term will be good enough to describe the charge conjugation odd part of the differential cross section.}
\end{abstract}

\maketitle

\date{}

\section{Introduction}
\label{introduction}
Following our previous paper \cite{Pire:2023kng}, we continue our investigation  of  the kinematical power suppressed (kinematical higher-twist) corrections  in the lowest order QCD amplitude of the reaction
\begin{equation}
 e^- e^+   \to M_1 M_2 \gamma\,,
\end{equation}
where the final state hadronic pair is now made of charged pseudoscalar mesons, typically  $\pi^+ \pi^-$ or $K^+ K^-$. The basic difference between the neutral meson case discussed in \cite{Pire:2023kng} and the present case is the presence of an initial state radiation (ISR) amplitude where the photon is emitted from the initial state lepton line. While such a QED process does not contribute to the production of a charge conjugation even ($C^+$) hadronic state (such as $\pi^0 \pi^0$), it turns out to be dominant \cite{Lu:2006ut} for the charge conjugation odd ($C^-$) hadronic state. Since we are interested in extracting the two meson  generalized distribution amplitudes (GDAs)  from experimental data, one needs to disentangle the QCD process from the ISR background, and this is possible by taking advantage of the different charge conjugation quantum number of the meson pair. A $C^-$odd observable such as an angular asymmetry is indeed proportional to the interference between $C^+$ and $C^-$ amplitudes, hence is linear in the GDA which enters linearly in the $C^+$ QCD amplitude.

The  ISR amplitude is entirely determined by QED and the experimental knowledge of meson timelike form factors, hence its validity is not restricted by any leading twist argument. On the other hand, the QCD amplitude factorizes \cite{Muller:1994ses} in the convolution of the GDA with a coefficient function, provided a leading twist dominance assumption  is used. Since feasible experiments are not at extremely large values of the large scale $Q^2$, it is thus of the utmost importance to quantify the power suppressed corrections  to the QCD amplitude, a task which has been made possible thanks to the breakthrough \cite{Braun:2011zr,Braun:2011dg,Braun:2011th, Braun:2022qly} in the understanding of target mass effects in the operator product expansion of two electromagnetic currents.
By  separating  kinematical and dynamical contributions in the product of two electromagnetic currents
$T \{j_{\mu}^\text{em}(z_1x)j_\nu^\text{em} (z_2x) \}$, they   proved that the kinematical corrections come from two types of operators, namely
the subtraction of traces in the leading-twist operators and the higher-twist operators which can be reduced to the total derivatives of the leading-twist ones.  They were then able to derive  gauge invariant and translation invariant amplitudes
for the deeply virtual Compton scattering (DVCS) process which take into account target mass and squared transferred momenta effects, without introducing new hadronic matrix elements. This technique was 
first applied to the DVCS reaction \cite{Braun:2012bg,Braun:2012hq, Braun:2014sta,Braun:2022qly}  in view of improving the extraction of the $t-$dependence of GPDs (generalized parton distributions) from experimental data, which is crucial to access the tomography of the nucleon \cite{Burkardt:2000za, Ralston:2001xs,Diehl:2002he}, then to the meson pair production in $\gamma^* \gamma$ collisions at $e^- e^+$ colliders \cite{Lorce:2022tiq, Lorce:2022cze} in view of improving the extraction of the $s-$dependence of GDAs from experimental data \cite{Pire:2002ut}.

The present paper is organized as follows. In Section~\ref{kinematics},  the kinematics of $e^- e^+   \to M \bar{M}  \gamma$ are discussed, and we give the complete formula for the interference term of the cross section which is expressed in terms of three helicity amplitudes.
A numerical estimate  for the cross section of this process is  provided in Section~\ref{fr}, and the kinematical power suppressed corrections are included up to $1/Q^2$ in this calculation. Our results are summarized in Sec.\,\ref{summary}.

\section{Kinematics and cross sections}
\label{kinematics}

\begin{figure}[htp]
\centering
\includegraphics[width=0.55\textwidth]{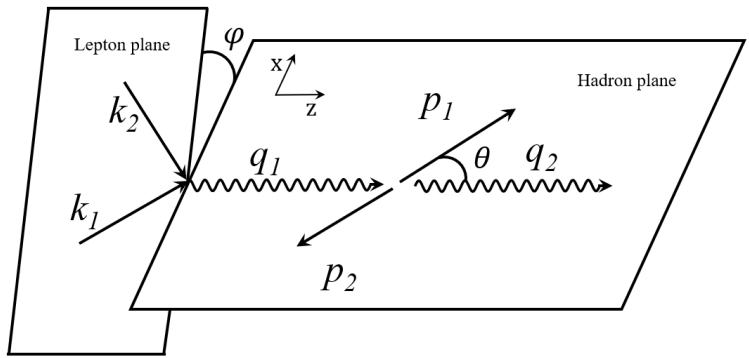}
\caption{Description of the kinematics of  the reaction  $e^-(k_1) e^+(k_2)  \to  \gamma^*(q_1) \to M(p_1) \bar{M}(p_2)  \gamma(q_2)$,  in the center-of-mass frame of the meson pair; the z axis is chosen along the photons momenta.  }
\label{fig:kin}
\end{figure}

In the process $e^- e^+  \to M \bar{M}  \gamma$, there are two types of subprocesses which differ by the charge conjugation of the meson pair.
The charge conjugation $C=+$ meson pair is produced in the subprocess (a): $e^- e^+  \to \gamma^* \to M \bar{M}  \gamma$, and the charge conjugation odd one is contributed by  the ISR  subprocess (b): $e^- e^+  \to \gamma^* \gamma \to M \bar{M}  \gamma$. To describe the reaction of $e^-(k_1) e^+(k_2)  \to  \gamma^*(q_1) \to M(p_1) \bar{M}(p_2)  \gamma(q_2)$,  the following variables are introduced,
\begin{align}
s=(k_1+k_2)^2, \qquad u=(k_1-q_2)^2, \qquad  \hat{s}=W^2=(p_1+p_2)^2, \qquad \beta_0=\sqrt{1-\frac{4 m^2}{\hat{s}}}
\label{eqn:kinva}
\end{align}
where $m$ is the meson mass. In Fig.\,\ref{fig:kin}, we show the  description of momenta and angles involved in this reaction.
The kinematics are discussed in the center-of-mass frame of the meson pair, and we choose a coordinate system with the $z$ axis along the momenta of the photons.
The momenta of the mesons lie in the x-z plane, and the $x$ component of $p_1$ is always positive,
\begin{align}
p_1=(p_1^0,\, |\bm{p_1}|\sin \theta,\, 0,\, |\bm{p_1}|\cos \theta),
\label{eqn:p1m}
\end{align}
where $\theta$ is the polar angle of $\bm{p_1}$, and it can be given in terms of Lorentz invariants,
\begin{align}
\cos{\theta}=\frac{q_1\cdot(p_2-p_1)}{\beta_0\,(q_1\cdot q_2)}.
\label{eqn:pola}
\end{align}
$\varphi$ is the azimuthal angle between the lepton plane and hadron plane as indicated in  Fig.\,\ref{fig:kin}, and it is expressed as
\begin{align}
\sin{\varphi}=\frac{   4  \epsilon_{\alpha \beta \gamma \delta}  q_1^{\alpha} q_2^{\beta}p_1^{\gamma} k_1^{\delta}  }{\beta_0 \sin{\theta} \sqrt{us \hat{s}(\hat{s}-u-s) }}
\label{eqn:amu}
\end{align}
where the convention $\epsilon_{0123}=1$ is used.
We introduce a parameter   $\zeta_0$ which indicates the longitudinal fraction of $\Delta=p_2-p_1$,
\begin{align}
\zeta_0= \frac{(p_2-p_1)\cdot q_2}{(p_2+p_1)\cdot q_2},
\label{eqn:skewness}
\end{align}
and one can obtain $\zeta_0=\beta_0 \cos{\theta}$ using Eq.\,(\ref{eqn:p1m}).
For convenience we define two lightlike vectors $n$ and $\tilde{n}$, and they can be expressed in terms of  $q_1$ and  $q_2$ as
\begin{align}
\tilde{n} = q_1-(1+\tau)q_2, \qquad  n=q_2
\label{eqn:lightlike}
\end{align}
with  $\tau=\hat{s}/(s-\hat{s})$. Combining with Eq.\,(\ref{eqn:skewness}), one obtains
\begin{align}
 \Delta=p_2-p_1=\zeta_0 (\tilde{n}-\tau n)+\Delta_T,  \qquad   2P=p_1+p_2=\tilde{n}+\tau n,
\label{eqn:vectors}
\end{align}
where $\Delta_T$ is the transverse component of $\Delta$, and it is given by
\begin{align}
\Delta_T^{\mu} =g_{\perp}^{\mu \nu}  \Delta_{\nu}
\label{eqn:dtrans}
\end{align}
with
\begin{align}
g_{\perp}^{\mu \nu} =g^{\mu\nu}-
\frac{n^{\mu}\tilde{n}^{\nu}+n^{\nu}\tilde{n}^{\mu}}{n\cdot \tilde{n}}.
\label{eqn:gtmu}
\end{align}
We can obtain $\Delta_T^2=g_{\perp}^{\mu \nu} \Delta_{\mu} \Delta_{\nu}=4m^2-(1-\zeta_0^2)\hat{s}$  by the on-shell condition.

\begin{figure}[htp]
\centering
\includegraphics[width=0.35\textwidth]{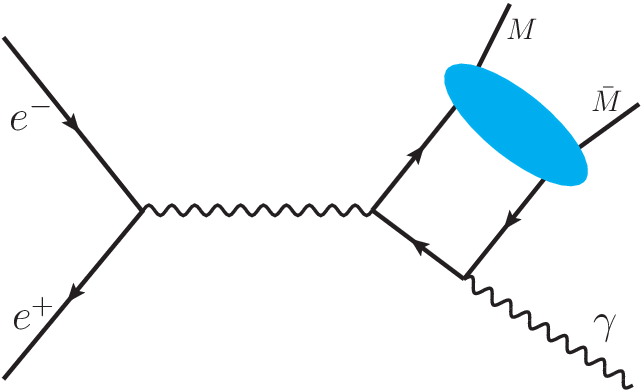}
\caption{
One of the leading-twist Feynman  diagrams for the subprocess (a): $e^- e^+   \to \gamma^{\ast} \to  M \bar{M} \gamma$;
the meson pair is produced with charge conjugation $C=+$ and the hadronization process is factorized with the help of 2 meson GDAs.}
\label{fig:gda}
\end{figure}

If $s=Q^2$ is large enough to satisfy the collinear QCD factorization \cite{Muller:1994ses} for the subprocess (a), the amplitude of $\gamma^* (Q^2) \to M \bar{M}  \gamma$ can be separated into a short-distance part $\gamma^* \to q \bar{q}  \gamma$ and  a long-distance part $  q \bar{q} \to M \bar{M}$ \cite{Lu:2006ut} as shown in Fig.\,\ref{fig:gda}.
The short-distance part is calculable at leading twist in perturbation theory, while the long-distance part  is described by a two-meson GDA \cite{Diehl:1998dk, Diehl:2000uv,Diehl:2003ny}. In case of a pseudoscalar meson pair, we define  the twist-2 GDA  $\Phi_q(z,\zeta_0, \hat{s})$ for the quark flavor $q$,
\begin{align}
\langle \bar{M}(p_2) M(p_1)  | \,\bar{q}(n/2) \slashed{n}q(-n/2)\, | 0 \rangle
=  2 P\cdot n \int_0^1 dz \,e^{i(2z-1) P\cdot n}\, \Phi_q(z,\zeta_0, \hat{s}),
\label{eqn:gdame}
\end{align}
where $z$ denotes  the momentum fraction of the quark, and the GDAs are actually dependent on a renormalization scale $\mu^2$ which is set  as $\mu^2=s$ here.
GDAs  evolve with the scale $Q^2$ according to  the Efremov-Radyushkin-Brodsky-Lepage evolution equation\cite{Lepage:1979zb}, and the general expressions of GDAs can be obtained by solving the evolution equation. For example, the charge conjugation $C=+$ GDAs are expanded in
Gegenbauer polynomials $C_n^{(3/2)}(x)$ and Legendre polynomials $P_l(x)$ \cite{Diehl:2000uv},
\begin{align}
\Phi_q(z, \cos \theta, \hat{s})=6\, z(1-z) \sum_{\substack{ n=1\\ n\,\text{odd} }}^{\infty} \sum_{\substack{l=0\\ l\,\text{even}}}^{n+1}
\tilde{B}_{nl}(\hat{s}) \,C_n^{(3/2)}(2z-1)\, P_l(\cos \theta),
\label{eqn:gda-expression}
\end{align}
where the subscript $l$ indicates the angular momentum of the  meson pair, and the scale dependence is abbreviated in $\tilde{B}_{nl}(\hat{s})$. In general, $\tilde{B}_{nl}(\hat{s})$ are complex functions,
\begin{align}
\tilde{B}_{nl}(\hat{s})=|\tilde{B}_{nl}(\hat{s})| e^{i*\delta_{l}},
\label{eqn:gda-im-ph}
\end{align}
where $i \delta_{l}$ are  imaginary phases.
If we consider the  asymptotic  limit of $Q^2 \rightarrow \infty$, only the  $n=1$  terms  remain in Eq.\,(\ref{eqn:gda-expression}) \cite{Diehl:2000uv},
\begin{align}
\Phi_q(z, \cos \theta, \hat{s})=18\, z(1-z) (2z-1) \left[\tilde{B}_{10}(\hat{s})
+\tilde{B}_{12}(\hat{s})  P_2(\cos \theta) \right],
\label{eqn:gda-exp-asm}
\end{align}
and they are the  asymptotic expressions of GDAs.

Let us introduce the following hadron tensor to discuss the amplitudes of  $\gamma^* \to M \bar{M}  \gamma$,
\begin{align}
A_{\mu \nu}=i\int d^4x\, e^{\frac{-i(q_1+q_2)\cdot x}{2}} \langle \bar{M}(p_2) M(p_1)  | \,
T \{ j_{\mu}^{\text{em}}(x/2)  j_\nu^{\text{em}} (-x/2) \} \, | 0 \rangle ; \! \!
\label{eqn:amp0}
\end{align}
it satisfies the electromagnetic gauge invariance condition $A_{\mu \nu}q_1^{\mu}=A_{\mu \nu}q_2^{\nu}=0$. Therefore, $A_{\mu \nu}$ can be expressed as \cite{Braun:2012bg}
\begin{align}
A^{\mu \nu}=-A^{(0)}\,g_{\perp}^{\mu\nu}+A^{(1)}\,
\left(\tilde n^\mu-(1+\tau)n^\mu\right) \frac{\Delta_T^{\nu}}{\sqrt{s}}
+\frac{1}{2}\, A^{(2)}\, \Delta_{\alpha}\Delta_{\beta}(g_{\perp}^{\alpha \mu} g_{\perp}^{\beta \nu}- \epsilon_{\perp}^{\alpha \mu}   \epsilon_{\perp}^{\beta \nu} )+ A^{(3) \mu}\, n^{\nu},
\label{eqn:amp1}
\end{align}
where the tensor $\epsilon_{\perp}^{ \mu \nu}$  is given by
\begin{align}
\epsilon_{\perp}^{\mu \nu} = \epsilon^{\mu \nu \alpha \beta}\,
 \frac{\tilde{n}_{\alpha} n_{\beta}}{n\cdot \tilde{n}}.
\label{eqn:gt}
\end{align}
$A^{(i)}  (i=0,1,2)$ denote three independent helicity amplitudes which are expressed in terms of GDAs, however, $A^{(3)}$ is not physical since it contributes only if  the outgoing real photon were longitudinally polarized.
In the center-of-mass frame of the meson pair, we follow the conventions of Ref.\,\cite{Diehl:2000uv} for the polarization vectors of the virtual photon,
\begin{align}
& \epsilon_{\pm}^{\mu}=\frac{1}{\sqrt{2}}(0, \mp 1, -i, 0), \quad
\epsilon_{0}^{\mu}=\frac{1}{\sqrt{s}}(|q_1^3|, 0, 0, q_1^0),
\label{eqn:pol-virt}
\end{align}
where $\epsilon_{\pm}$ are the transverse polarization vectors, and  $\epsilon_0$ indicates longitudinal  polarization. Only transverse polarization vectors  $\tilde{\epsilon}_{\pm}$ exist for the real photon, and they are same as the corresponding ones of the virtual photon, $\tilde{\epsilon}_{\pm}=\epsilon_{\pm}$.
The helicity amplitudes are defined as  $A_{i j}=  \epsilon_{i}^{\mu}   \tilde{\epsilon}_{j}^{\ast \nu } A_{\mu \nu}$, and they are expressed in terms of $A^{(i)}$,
\begin{align}
A_{++}=A_{--}=A^{(0)}, \quad \quad
A_{0 \pm }=-A^{(1)} (\Delta \cdot \epsilon_{\mp}), \quad \quad
A_{\pm \mp}=-A^{(2)} (\Delta \cdot \epsilon_{\pm})^2.
\label{eqn:am-hel-ind}
\end{align}
If we limit our study to a leading order in $\alpha_s$ calculation, the amplitudes $A_{0 \pm }$ and $A_{\pm \mp}$ are respectively of order $1/Q$ and $1/Q^2$ effects ($s=Q^2$), and only $A_{++}$  receives a leading $1/Q^0$ contribution which has been calculated in Ref.\,\cite{Lu:2006ut}. Then, the differential cross section of $e^- e^+ \to M \bar{M} \gamma$ reads \cite{Pire:2023kng}
\begin{align}
\frac{d \sigma_{\text{G}}}{d\hat{s} \,du\, d(\cos \theta)\, d\varphi}=&
\frac{\alpha_{\text{em}}^3 \beta_0}{16 \pi s^3 } \,\frac{1}{1+\epsilon}  \,\Big[ |A_{++}|^2+ |A_{-+}|^2+2\epsilon\, |A_{0+}|^2 -2 \text {sgn}( \rho)\sqrt{\epsilon(1-\epsilon)} \nonumber \\
&\times  \text{Re}(A_{++}^{\ast} A_{0+} -A_{-+}^{\ast} A_{0+}) \cos \varphi  + 2 \epsilon \, \text{Re}(A_{++}^{\ast} A_{-+}) \cos (2\varphi) \Big], \nonumber
\end{align}
where the subscript $G$ indicates that the meson pair is produced through the subprocess (a), namely, in the charge conjugation $C=+$ state, may be rewritten neglecting terms of order $1/Q^3$ and higher as
\begin{align}
\frac{d \sigma_{\text{G}}}{d\hat{s} \,du\, d(\cos \theta)\, d\varphi}=&
\frac{\alpha_{\text{em}}^3 \beta_0}{16 \pi s^3 } \,\frac{1}{1+\epsilon}  \,\Big[ |A_{++}|^2+2\epsilon\, |A_{0+}|^2 -2 \text {sgn}( \rho)\sqrt{\epsilon(1-\epsilon)} \nonumber \\
&\times  \text{Re}(A_{++}^{\ast} A_{0+} ) \cos \varphi  + 2 \epsilon \, \text{Re}(A_{++}^{\ast} A_{-+}) \cos (2\varphi) \Big] .
\label{eqn:epho-cro}
\end{align}
$\epsilon$ is the usual polarization parameter which is written in terms of  Lorentz invariants as:
\begin{align}
\epsilon=\frac{y-1}{1-y+\frac{y^2}{2}}, \qquad y=\frac{q_1\cdot q_2}{k_1\cdot q_2},
\label{eqn:polar}
\end{align}
and  $\text {sgn}(\rho) =  |\rho|/\rho$ is the sign function with $\rho=\hat{s}-s-2u$.
Note that Eq.~(\ref{eqn:epho-cro}) is different from the cross section presented in Ref.\,\cite{Lu:2006ut}, where only the leading-twist amplitude $A_{++}$ is included and the variable  $y_{\text{Lu}}$ is used instead of $u$,
\begin{align}
y_{\text{Lu}}=1+\frac{u}{s-\hat{s}}.
\label{eqn:utoy}
\end{align}

Recently, the authors of Refs.\,\cite{Braun:2011zr,Braun:2011dg,Braun:2011th} separated the  kinematical contribution from the dynamical one in the operator product of two electromagnetic currents $T \{j_{\mu}^\text{em}(z_1x)j_\nu^\text{em} (z_2x) \}$ up to twist-4 accuracy,
and the kinematical contribution does not contain the genuine higher-twist operators, namely it is expressed in terms of the subtraction of traces in the leading-twist operators and the total derivatives of the leading-twist operators. If one includes the kinematical corrections to the cross sections of the off-forward hard reactions  where GPDs or GDAs can be accessed, hopefully more precise cross sections can be obtained with the leading-twist GPDs or GDAs.
In Ref.\,\cite{Pire:2023kng}, we derived that the up to twist 4  kinematical power suppressed corrections  are  included in the helicity amplitudes of   $e^- e^+  \to M \bar{M}  \gamma$  as
\begin{align}
A^{(0)}&= \chi \left\{  \left(1+ \frac{\hat{s}}{2s}\right) \int_0^1 dz\, \frac{\Phi(z, \eta, \hat{s})}{1-z}
+\frac{\hat{s}}{s}\int_0^1  dz \, \frac{\Phi(z,\eta, \hat{s})}{z}\, \ln(1-z)
\right. \nonumber \\
&\qquad +\left. \left(\frac{2\hat{s}}{s} \,\eta   +\frac{\Delta_T^2}{\beta_0^2 s} \frac{\partial}{\partial \eta} \right)
\frac{\partial}{\partial \eta}
  \int_0^1 dz \,\frac{ \Phi(z,\eta, \hat{s}) }{z} \left[ \frac{\ln(1-z)}{2} +\text{Li}_2(1-z) -\text{Li}_2(1)  \right]
  \right \}, \nonumber \\
A^{(1)}&=  \frac{2\chi}{\beta_0 \sqrt{s}} \frac{\partial}{\partial  \eta }   \int_0^1 dz \, \Phi(z, \eta, \hat{s}) \,
\frac{\ln(1-z)}{z}, \nonumber \\
A^{(2)}&=\frac{2 \chi}{ \beta_0^2 s} \frac{\partial^2}{\partial \eta^2}       \int_0^1 dz  \,\Phi(z, \eta, \hat{s}) \,
\frac{2z-1}{z}\, \ln(1-z),
\label{eqn:int-gdas-a}
\end{align}
where  $\eta=\cos \theta$,  and $\chi=5e^2/18$.
$\Phi=\Phi_u+\Phi_d$ is defined  for the isosinglet pion meson pair.
Note that one needs to replace  $\chi \Phi$ by
$e_u^2 \Phi_u+e_d^2 \Phi_d +e_s^2 \Phi_s$ in the helicity amplitudes if a $K$ meson pair is produced in the charge conjugation $C=+$ state, since
the GDA for $s$ quark should be included.
There are two types of kinematical corrections $\mathcal O(\hat{s}/s)$ and $\mathcal O(\Delta_T^2/ s)$ in the helicity amplitudes, and the latter one also contains the so-called target mass correction $\mathcal O(m^2/s)$ using $\Delta_T^2=4m^2-(1-\zeta_0^2) \hat{s}$.
Eqs.\,(\ref{eqn:gda-expression}) and (\ref{eqn:int-gdas-a}) imply that
the $S$-wave GDAs will not contribute to the helicity flipped amplitudes $A_{0 \pm }$ and $A_{\pm \mp}$.

\begin{figure}[htp]
\centering
\includegraphics[width=0.55\textwidth]{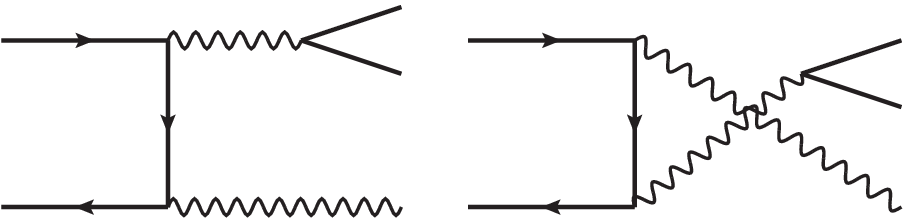}
\caption{
Feynman  diagrams of the initial state radiation (ISR) subprocess for the reaction $e^- e^+  \to M \bar{M} \gamma$;  the meson pair is produced with charge conjugation $C=-$ and the hadronic process occurs through the meson timelike form factors.}
\label{fig:cha}
\end{figure}

The charge conjugation $C=-$ meson pair is produced by ISR in the reaction $e^- e^+   \to M \bar{M} \gamma$, and the corresponding Feynman  diagrams are shown in Fig.\,\ref{fig:cha}.
Note that the amplitude of this process is not described in terms of a twist expansion, since the hadronic part entirely comes from the timelike electromagnetic  meson form factor $F_M(\hat s)$, which is experimentally known at moderate values of $\hat s$ and does not need to be calculated from QCD (which indeed would be difficult in this nonperturbative region).
The differential cross section for ISR is given in Refs.\,\cite{Lu:2006ut,BaBar:2015onb,Arbuzov:1997je},
\begin{align}
\frac{d \sigma_{\text{ISR}}}{d\hat{s} \,du\, d(\cos \theta)\, d\varphi} = &
\frac{\alpha_{\text{em}}^3 \beta_0^3}{8 \pi s^2}\,\frac{|F_M(\hat{s})|^2}{\epsilon \hat{s}}   \,\Big[ (1-2x(1-x))\sin^2\theta + 4x(x-1) \epsilon \cos^2\theta
+\text{sgn}(\rho)(2x-1)\sqrt{2x(x-1)}\sqrt{\epsilon(1-\epsilon)} \nonumber \\
 &\times  \sin(2\theta) \cos \varphi    +2 x(1-x) \epsilon \sin^2\theta    \cos (2\varphi)  \Big],
\label{eqn:cro-isr}
\end{align}
where the variable $x$ is defined as
\begin{align}
x=\frac{(q_1)^2}{2q_1\cdot q_2}=\frac{s}{s-\hat{s}}.
\label{eqn:polarx}
\end{align}

There is also an interference  between the  two subprocesses amplitudes in the cross section, and it can be expressed by
\begin{align}
\frac{d \sigma_{\text{I}}}{d\hat{s} \,du\, d(\cos \theta)\, d\varphi} = &
\frac{\alpha_{\text{em}}^3 \beta_0}{8 \pi s^2}\,\frac{\sqrt{2} \beta_0 }{\sqrt{\hat{s} s \epsilon(1+\epsilon) }  }   \,\Big[  C_0 +C_1 \cos \varphi +C_2 \cos (2\varphi) +C_3 \cos (3\varphi)     \Big],
\label{eqn:cro-int}
\end{align}
where the coefficients $C_i$ read
\begin{align}
C_0=&-\text{sgn}(\rho)\,\sqrt{\epsilon(1-\epsilon)}  \sqrt{2x(x-1)} \text{Re}(A_{++}F_M^{\ast})\cos \theta+\text{sgn}(\rho)\,(x-1)\sqrt{\epsilon(1-\epsilon)}
\text{Re}(A_{0+}F_M^{\ast}) \sin \theta,\nonumber \\
C_1=&- \left [1-(1-x)(1-\epsilon)\right ]  \text{Re}(A_{++}F_M^{\ast}) \sin \theta +2  \epsilon  \sqrt{2x(x-1)} \text{Re}(A_{0+}F_M^{\ast})  \cos \theta
+(x-1) \text{Re}(A_{-+}F_M^{\ast})  \sin \theta, \nonumber \\
C_2=&\text{sgn}(\rho)\, \sqrt{\epsilon(1-\epsilon)} x \text{Re}(A_{0+}F_M^{\ast})\sin \theta +\text{sgn}(\rho)\, \sqrt{\epsilon(1-\epsilon)} \sqrt{2x(x-1)}\text{Re}(A_{-+}F_M^{\ast}) \cos \theta, \nonumber\\
C_3=&- \epsilon x \text{Re}(A_{-+}F_M^{\ast})   \sin \theta.
\label{eqn:int-co}
\end{align}

If one compares Eqs.\,(\ref{eqn:epho-cro}) and (\ref{eqn:cro-isr}) with (\ref{eqn:cro-int}), the following relation can be found,
\begin{align}
d \sigma_{\text{G}} :d \sigma_{\text{I}}: d \sigma_{\text{ISR}} \sim  \frac{1}{s}:\frac{1}{\sqrt {s \hat s}} : \frac{1}{\hat s},
\label{eqn:cro-3}
\end{align}
and this is reminiscent of the different $s$ and $Q^2$ behavior of the Bethe Heitler and QCD amplitudes for the DVCS or timelike Compton scattering~\cite{Berger:2001xd} processes.
Due to the factorization condition $s \gg \hat{s}$, the largest contribution comes from the ISR  cross section  $d \sigma_{\text{ISR}}$. However, what we are interested in are the GDAs, which do not contribute to the ISR  cross section of Eq.\,(\ref{eqn:cro-isr}). To access the GDAs, it is thus better to measure the interference cross section.
Besides,  the imaginary phases of GDAs cannot be extracted from Eq.\,(\ref{eqn:epho-cro}), and we can illustrate this  using the asymptotic GDAs of Eq.\,(\ref{eqn:gda-exp-asm}).
 Since only the $D$-wave GDAs contribute to the helicity flipped amplitudes $A_{0 \pm }$ and $A_{\pm \mp}$ as indicated by Eq.\,(\ref{eqn:int-gdas-a}),
the term of $\text{Re}(A_{-+}^{\ast} A_{0+})$  in Eq.\,(\ref{eqn:epho-cro}) should be independent on the imaginary phases of GDAs, and the  phases affect the cross section via  $\text{Re}(\tilde{B}_{10}(\hat{s}) \tilde{B}_{12}^{\ast}(\hat{s}) )$ for the remaining terms of Eq.\,(\ref{eqn:epho-cro}), thus only the relative phase between $\tilde{B}_{10}(\hat{s})$ and  $\tilde{B}_{12}(\hat{s})$ can be determined for the GDAs. On the contrary, the complete information of GDAs can be accessed by analyzing the interference cross section due to  $d \sigma_{\text{I}}\propto  \text{Re}( A_{ij}F_M^{\ast}(\hat{s}))$.

The hadron GDAs are extracted from  $e^- e^+  \to \gamma^* \to M_1 M_2  \gamma$, where electromagnetic interaction and strong interaction are involved.
The first moments of GDAs lead to the enegy-momentum tensor (EMT) FFs of hadrons, and  the latter ones are difficult to be accessed by experiment directly due  to the weak gravitational interaction. Thus, the hadron GDAs are a powerful way to  obtain the timelike EMT FFs.  The imaginary phases are used when one transfers the timelike FFs to the  spacelike ones~\cite{Strikman:2010pu,Miller:2010tz,Klopot:2013laa},
\begin{align}
F^h (t) & = \int_{4 m_h^2}^\infty \frac{d\hat{s}}{\pi}
            \frac{{\rm Im}\, F^h (\hat{s})}{\hat{s}-t},
\label{eqn:dispersion-form-1}
\end{align}
where  $F^h (\hat{s})$ is a EMT or electromagnetic FF for the hadron, and $m_h$ is the hadron mass.
Moreover, many interesting physical quantities of hadrons can be gained from the spacelike EMT FFs, such as mass radius, pressure and shear force distributions~\cite{Polyakov:2018zvc, Burkert:2018bqq,Burkert:2023wzr,Lorce:2018egm,Kumericki:2019ddg,Dutrieux:2021nlz,Kumano:2017lhr,Wang:2021ujy,Raya:2021zrz,Freese:2019bhb,Chakrabarti:2020kdc}. As a consequence, the measurement of the interference cross section will play an important role in the GDA extraction. Since the charge conjugation $C=+$  and $C=-$ meson pairs are produced in the subprocesses (a) and (b), respectively, Eqs.\,(\ref{eqn:epho-cro}) and (\ref{eqn:cro-isr}) will remain the same if one interchanges $M$ and $\bar{M}$, but the interference cross section of Eq.\,(\ref{eqn:cro-int}) will change its sign under this interchange, so that only the interference term survives in $d \sigma(M\bar{M})-d \sigma(\bar{M}M)=2d \sigma_{\text{I}} $, and it has been measured by BaBar for $\pi^+ \pi^-$ \cite{BaBar:2015onb}.

\section{Numerical estimates of  the cross sections}
\label{fr}

In the following we present our numerical estimates for the various components of the cross section of $e^- e^+ \to \pi^+ \pi^- \gamma$ by using the formulas derived in Section \ref{kinematics} and we show the importance of the kinematical power suppressed corrections on various observables. We first calculate the differential cross section emerging from the dominant ISR process for BESIII and Belle II kinematics. We then evaluate in these kinematics both the pure QCD and the QCD-ISR interference contributions to differential cross sections, as a function of the hadron pair invariant mass, and as a function of the azimuthal angle. We use two models for the $\pi \pi$ GDA, namely one coming from a previous analysis \cite{Kumano:2017lhr} of Belle experimental data then a model based on the asymptotic shape predicted by QCD evolution of GDAs.

\subsection{Cross section of the ISR process}
\begin{figure}[htp]
\centering
\includegraphics[width=0.85\textwidth]{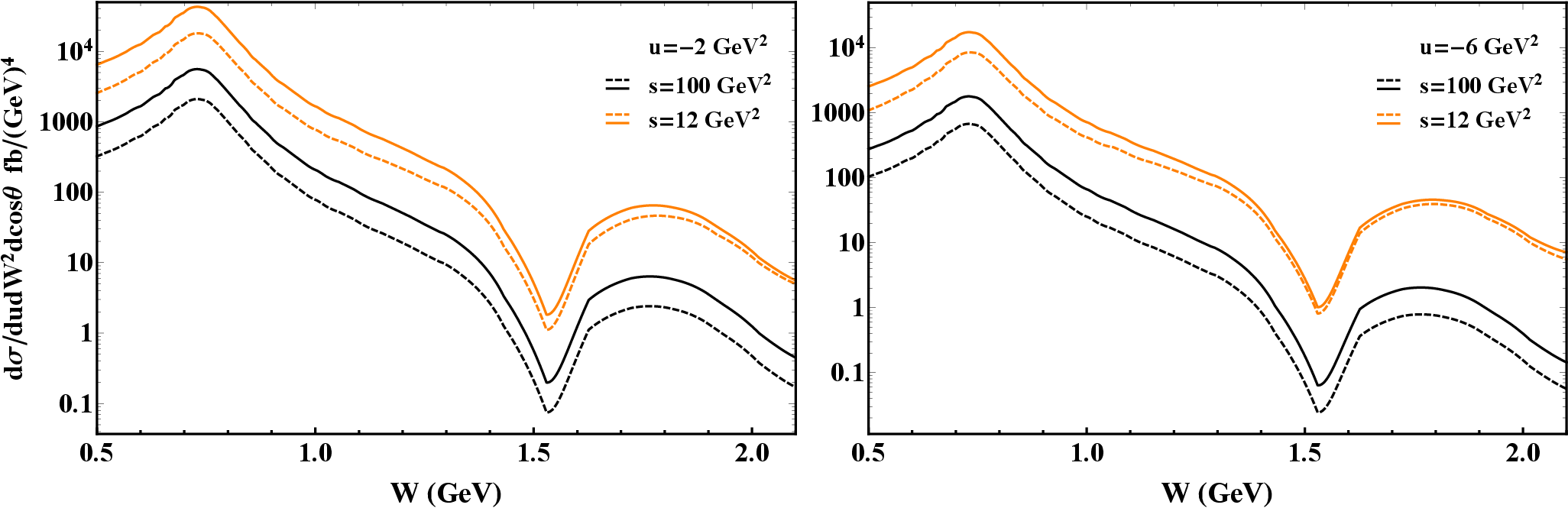}
\caption{The differential cross section $d\sigma/du dW^2 dcos\theta$ for the ISR process; $s$ is set as $s=12 $ GeV$^2$ for BESIII and $s=100 $ GeV$^2$ for Belle II.
We choose $u=-2 $ GeV$^2$ and  $u=-6 $ GeV$^2$ for the left and right panels, respectively.
The solid lines denote $\cos \theta=0.2$, and the dashed lines represent $\cos \theta=0.8$.}
\label{fig:num1}
\end{figure}

As indicated by Eq.~(\ref{eqn:cro-3}), the ISR process  gives the dominant contribution to the cross section.
In this calculation, the absolute value of the timelike electromagnetic form factor $|F_{\pi}(\hat{s})|$ is needed, and we use the parametrization presented in Ref.~\cite{Bartos:2023uch}, where the pion form factor was extracted from experimental measurements of $e^- e^+  \to  \pi^+ \pi^- (\gamma)$  \cite{BaBar:2012bdw,BESIII:2015equ,Xiao:2017dqv}.
Integrating Eq.~(\ref{eqn:cro-isr}) over $\varphi$, we present the cross section for the ISR process in Fig.~\ref{fig:num1}.
We choose $s=12 $ GeV$^2$ and $s=100 $ GeV$^2$ which are typical values for BESIII and Belle II, respectively. $u$ is set as $u=-2 $ GeV$^2$ in the left panel and $u=-6 $ GeV$^2$ in the right panel, and the solid (dashed) curve denotes $\cos \theta=0.2$(0.8).
The predicted cross section of BESIII is about 10 times larger than the one of Belle II, and both of them decrease rapidly as $W$ increases from 0.7 GeV to 2.1 GeV.
In the figure, the peaks around $W \sim 0.7$ GeV and $W \sim 1.8$ GeV exist due to the oscillating form factor, and the first one comes from the  $\rho$ meson resonance.
Note that the differential cross section of the ISR process will increase as the emitted photon ($q_2$) becomes parallel to the momenta of the electron or positron, namely when $u=0$ or $u=W^2-s$ is satisfied. One can see that the cross sections in the left panel are larger than those of the right panel, since the momentum of  the emitted photon is closer to  the parallel direction at $u=-2 $ GeV$^2$.

\subsection{Cross sections  $d \sigma_{\text{I}}$ and $d \sigma_{\text{G}}$ with the extracted $\pi\pi$ GDA}

\begin{figure}[htp]
\centering
\includegraphics[width=0.5\textwidth]{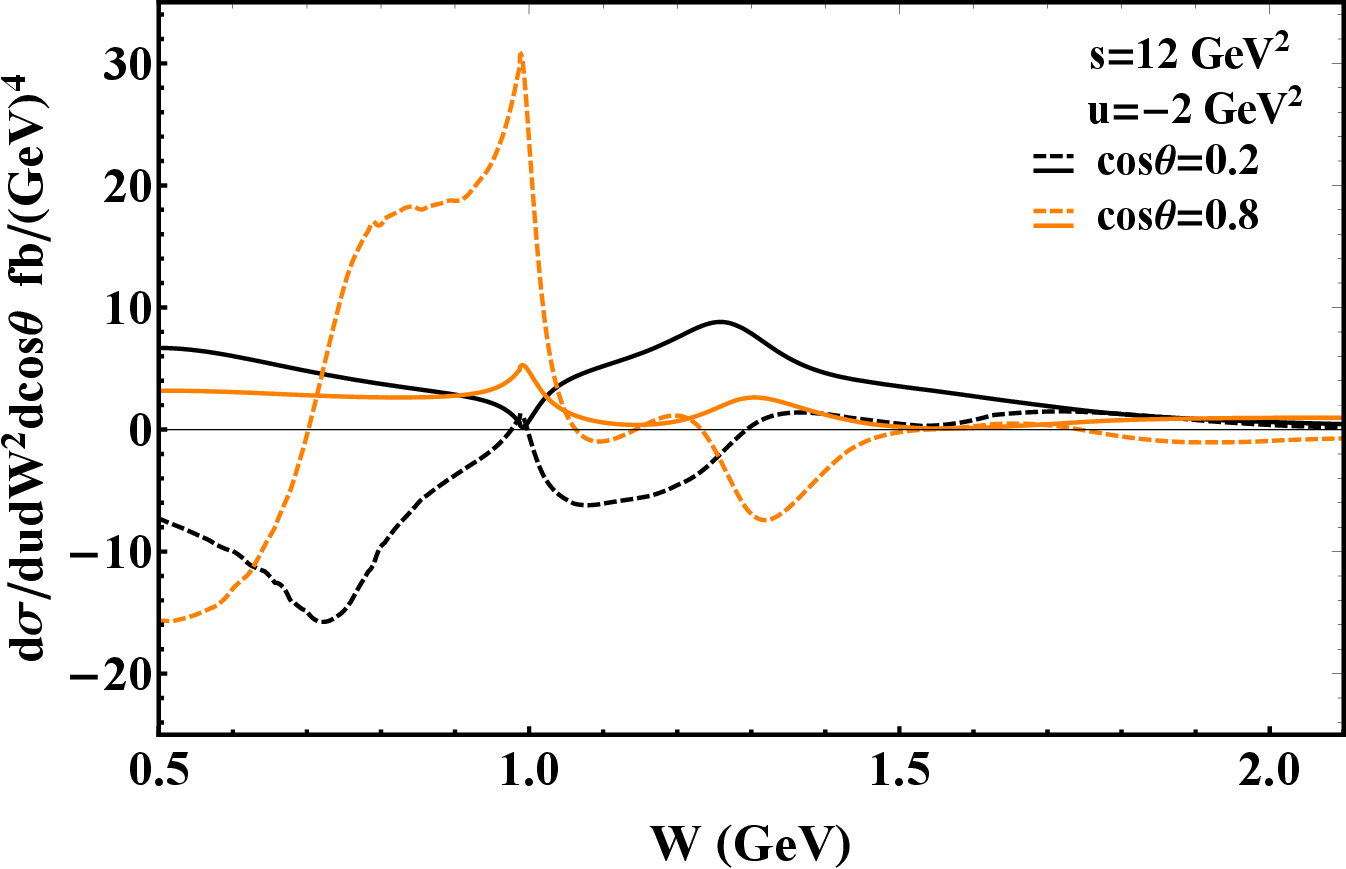}
\caption{ The meson pair invariant mass ($W$) dependence of the cross sections  $d \sigma_{\text{G}}$ and $d \sigma_{\text{I}}$ at $s=12 $ GeV$^2$; the model for the $\pi \pi$ GDA  \cite{Kumano:2017lhr} comes from an analysis of Belle measurements. $d \sigma_{\text{G}}$ is depicted as solid curves, and the dashed curves represent the interference term $d \sigma_{\text{I}}$.}
\label{fig:num2}
\end{figure}

 Since we are interested in extracting the GDAs, which appear in the other two contributions to the cross section for $e^- e^+  \to  \pi^+ \pi^- \gamma$, we also show the numerical results  for $d \sigma_{\text{G}}$ and the interference term  $d \sigma_{\text{I}}$. Integrating Eqs.~(\ref{eqn:epho-cro}) and (\ref{eqn:cro-int}) over $\varphi$, one  obtains:
\begin{align}
\frac{d \sigma_G}{du\, dW^2\, d(\cos \theta)}=&
\frac{\alpha_{\text{em}}^3 \beta_0}{8s^3 } \,\frac{1}{1+\epsilon}   \,\Big[ |A_{++}|^2+ |A_{-+}|^2+2\epsilon\, |A_{0+}|^2  \Big],\label{eqn:croig1}  \\
\frac{d \sigma_{\text{I}}}{du\, dW^2 \,d(\cos \theta)}=&
\frac{\alpha_{\text{em}}^3 \beta_0}{4 s^2}\,\frac{\sqrt{2} \beta_0 }{\sqrt{\hat{s} s \epsilon(1+\epsilon) }  }   C_0\,,
\label{eqn:croig}
\end{align}
where the meson GDA is required to calculate the helicity amplitudes  $A_{i j}$, and the electromagnetic form factor is also needed for the interference term $d \sigma_{\text{I}}$ in addition to the GDA.
In this work, we use two types of GDAs for the pion meson, one of them was extracted  from Belle measurements of $\gamma^{\ast} \gamma \to \pi^0 \pi^0$~\cite{Kumano:2017lhr}, the other is called the asymptotic $\pi \pi$ GDA~\cite{Diehl:2000uv}. As for the timelike form factor, only the absolute values are provided in Ref.~\cite{Bartos:2023uch}, and the imaginary phase of the form factor can be related to the phase shift of the $P$-wave $\pi \pi$ elastic scattering below the $KK$ threshold, however, the imaginary phase is unknown  above the $KK$ threshold.  The main purpose of this work is to check whether  the kinematical power suppressed  effect is important or not in the interference term, namely, the proportion of  power suppressed  contribution to the cross section of $d \sigma_{\text{I}}$,  which is not affected by the form factor since it is just an overall function as indicated by Eq.~(\ref{eqn:int-co}). Therefore, the absolute value of the meson form factor is used in calculating the interference term.

Using the extracted $\pi \pi$ GDA, we include the   kinematical power suppressed corrections to the cross sections  $d \sigma_{\text{G}}$ and $d \sigma_{\text{I}}$ according to Eq.~(\ref{eqn:int-gdas-a}), and
they are presented in Fig.~\ref{fig:num2} with $s=12 $ GeV$^2$ and $u=-2 $ GeV$^2$. We choose
 $\cos \theta=0.2$ and  $\cos \theta=0.8$ as two typical values for the polar angle.
The solid lines denote the cross section of  $d \sigma_{\text{G}}$, and the dashed curves indicate the interference term $d \sigma_{\text{I}}$. We can see that both $d \sigma_{\text{G}}$ and $d \sigma_{\text{I}}$ are much smaller than the cross section of the ISR process shown in Fig.~\ref{fig:num1}.
 The interference term $d \sigma_{\text{I}}$ is a few times larger than $d \sigma_{\text{G}}$, which is consistent with the prediction of Eq.~(\ref{eqn:cro-3}).
 On the one hand, $d \sigma_{\text{I}}$ is much easier to  access in an experiment due to the larger cross section which can be obtained by taking the difference between the cross section of $e^- e^+  \to  \pi^+ \pi^- \gamma$ and the one of the same process where the $\pi^+$ and $\pi^-$ momenta are interchanged. On the other hand, the imaginary phases of GDAs cannot be fully determined by analyzing $d \sigma_{\text{G}}$, since
  the helicity amplitude squared in Eq.~(\ref{eqn:croig1}) are only dependent on the relative phases, however, they can be extracted from $d \sigma_{\text{I}}$ which are necessary to investigate the spacelike EMT FFs of hadrons as indicated by Eq.~(\ref{eqn:dispersion-form-1}).

\begin{figure}[htp]
\centering
\includegraphics[width=0.85\textwidth]{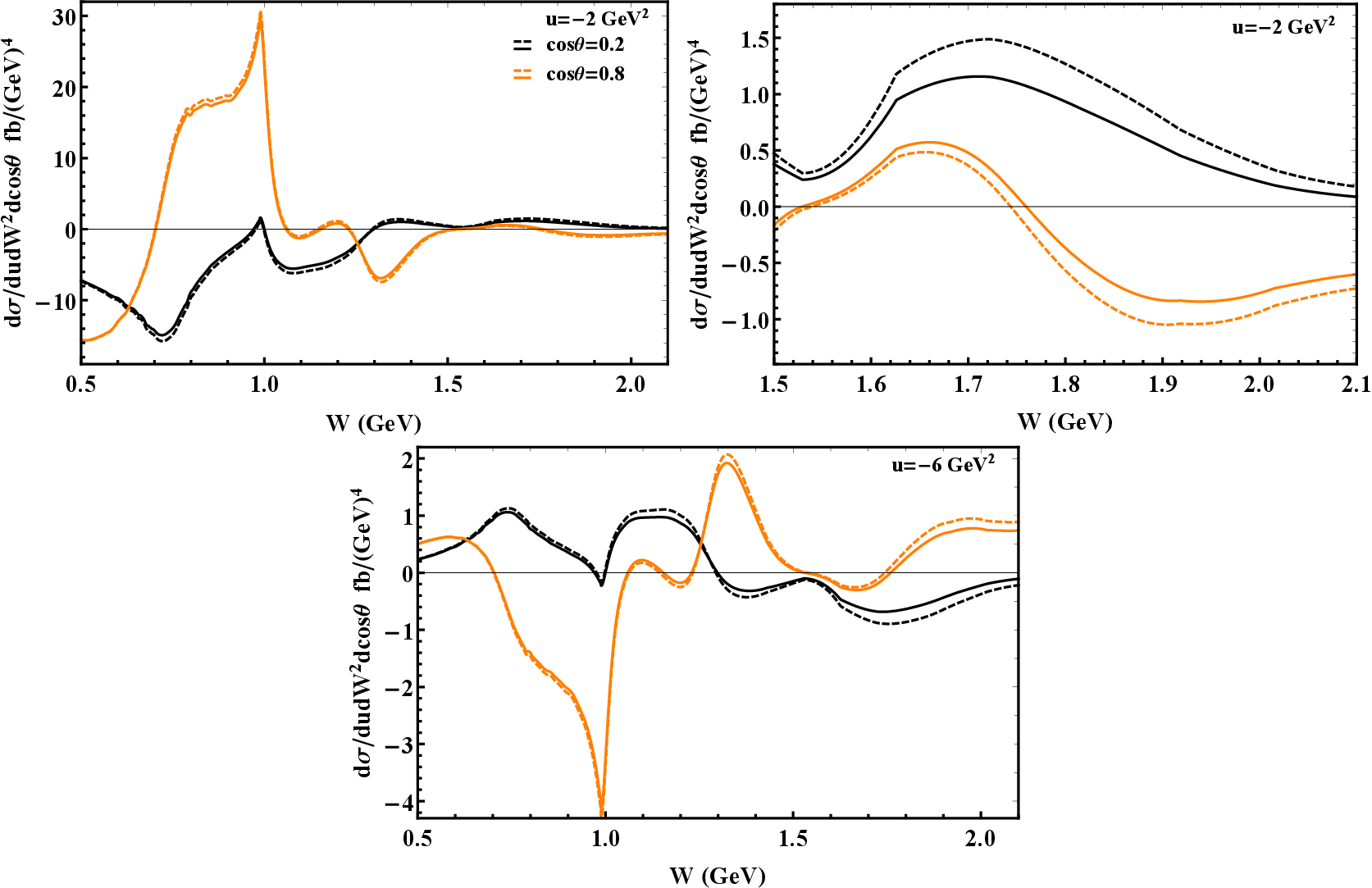}
\caption{Differential cross section of the interference term  $d \sigma_{\text{I}}$. The  kinematics is set  according to the BESIII experiment as $s=12 $ GeV$^2$,  $u=-2 $ GeV$^2$($-6 $ GeV$^2$) together with $\cos \theta=0.2$ (0.8). The dashed lines are the twist-2 cross
sections, and the solid lines include the  kinematical power suppressed contribution. The GDA model used comes from the estimate \cite{Kumano:2017lhr}  using the Belle measurements. }
\label{fig:num3}
\end{figure}

Since the interference term  $d \sigma_{\text{I}}$ plays an important role in the extraction of GDAs, it is necessary to use a precise description for  $d \sigma_{\text{I}}$. We employ the extracted $\pi \pi$ GDA to check the effect of  kinematical power suppressed corrections to the cross section. In Fig.~\ref{fig:num3}, the dashed curves represent the twist-2 contribution to  the interference term  $d \sigma_{\text{I}}$, and the solid ones indicate the cross sections with  the kinematical power suppressed corrections included up to twist 4. We choose  $s=12 $ GeV$^2$ which is a typical value for BESIII, and $u$ is set as  $u=$-2 GeV$^2$ and $u=$-6 GeV$^2$ for the top panel and bottom panel, respectively.  The back lines denote $\cos \theta=0.2$, and the orange ones represent $\cos \theta=0.8$ as indicated by the figure.
In the top left panel, the gaps between the solid lines and dashed lines are very small compared with the cross section if $W<1$ GeV, and the kinematical power suppressed corrections can account for $10-40\%$ of the cross section for the region of $W>1$ GeV. Since the cross sections decrease rapidly, and the gaps cannot clearly be seen in the region of $W>1.5$ GeV, the cross sections are also  depicted with 1.5 GeV $  \leq  W \leq $ 2.1 GeV in the top right panel. Similarly, one can infer that the  kinematical power suppressed corrections cannot be neglected when $W>1$ GeV from the bottom panel, where $u=$-6 GeV$^2$ is chosen.
As a consequence, it is necessary to  include the kinematical contribution to extract the pion GDA from  the interference term  $d \sigma_{\text{I}}$ measured at BESIII, since the kinematical contribution
is important to describe the cross section at large $W$ region. The timelike EMT FFs are obtained from the meson GDAs, and  timelike FFs can be transferred to the spacelike ones by the integral of
Eq.~(\ref{eqn:dispersion-form-1}). Since timelike EMT FFs decrease as the  $W=\sqrt{\hat{s}}$ goes up, one can usually set  $\hat{s} \sim 4 $ GeV$^2$ as the upper limit of the integral~\cite{Kumano:2017lhr}, which means that the timelike FFs at large $\hat{s}$ region (1 GeV$^2$$<\hat{s}=W^2<4$ GeV$^2$) are needed to make the integral convergent.
Thus, it is crucial to include the kinematical power suppressed corrections so as to extract the GDAs in the medium to large $W$ region precisely, which then can be used to study the  EMT FFs for mesons.

\begin{figure}[htp]
\centering
\includegraphics[width=0.5\textwidth]{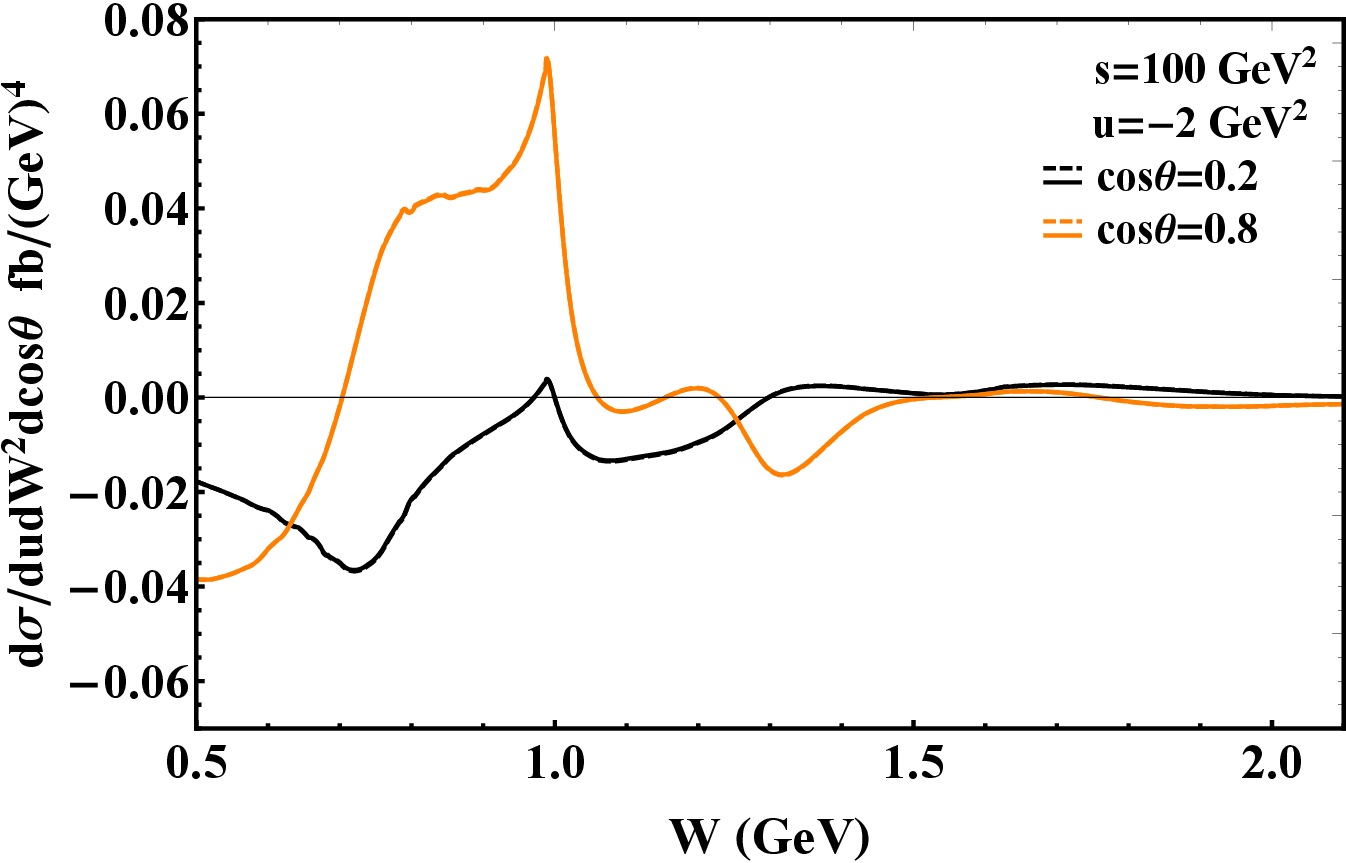}
\caption{Differential interference cross section   $d \sigma_{\text{I}}$;  $s=100 $ GeV$^2$ is chosen according to the  Belle II experiment kinematics.
The dashed curves denote the twist-2 cross sections, and the solid ones include  the kinematical power suppressed contribution.}
\label{fig:num4}
\end{figure}

If we increase $s$ from $12 $ GeV$^2$  to $100 $ GeV$^2$, which is typical for Belle II, in Fig.~\ref{fig:num4} the dashed  and  solid curves indicate  the twist-2 cross sections and the cross sections  with the kinematical power suppressed contribution included for the interference term, respectively. We choose  $\cos \theta=0.2$, $\cos \theta=0.8$ and $u=$-2 GeV$^2$ as shown in the figure.
One can predict that the interference term will diminish as $s$ goes up due to the factor $1/(s^2 \sqrt{s})$ from Eq.\,(\ref{eqn:croig}).
Compared with Fig.~\ref{fig:num3}, the cross sections become quite tiny when $s=100 $ GeV$^2$.
In Fig.~\ref{fig:num4}, we can hardly see the difference between the dashed and solid lines, showing  that the kinematical power suppressed corrections can be neglected for the interference term of $e^- e^+  \to  \pi^+ \pi^- \gamma$ at Belle II.  This also conforms to what is expected, since the kinematical correction is proportional to $\hat{s}/s$, and it only accounts for  $\sim 1\%$ of the cross section  in the kinematics of the measurements at Belle II.

\begin{figure}[htp]
\centering
\includegraphics[width=0.85\textwidth]{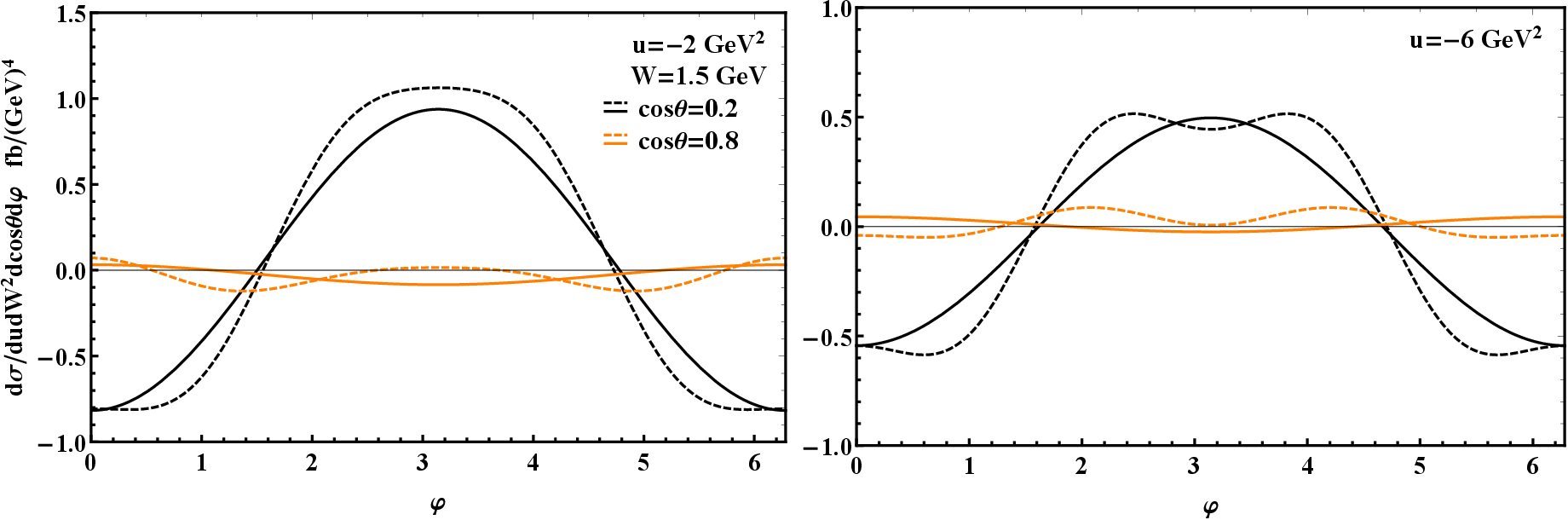}
\caption{The azimuthal angle $\varphi$ dependence of the interference cross section  $d \sigma_{\text{I}}$;  $s=12 $ GeV$^2$ and the extracted GDA is used. The dashed curves denote the twist-2 cross sections, and the solid ones include the kinematical power suppressed contribution.}
\label{fig:num4a}
\end{figure}

We have calculated the interference term  $d \sigma_{\text{I}}$ using Eq.\,(\ref{eqn:croig}), where the azimuthal angle $\varphi$ is integrated. In Eq.\,(\ref{eqn:croig}), only helicity amplitudes $A_{++}$ and $A_{0+}$ contribute to the interference term, however,  all three helicity amplitudes appear in the $\varphi$-dependent differential cross section of Eq.\,(\ref{eqn:cro-int}). Actually, the $\varphi$-dependent differential cross section of $d \sigma_{\text{I}}$ has been measured by the Babar collaboration~\cite{BaBar:2015onb}.
In order to check how the kinematical corrections affect the  $\varphi$-dependence of the interference term, we also estimate the  cross section $d \sigma_{\text{I}}/d\varphi$ with the help of  Eq.\,(\ref{eqn:cro-int}).
In Fig.\,\ref{fig:num4a}, the solid (dashed) lines show the cross sections without (with)  the kinematical power suppressed corrections included.
$s=12 $ GeV$^2$ and $W=1.5$ GeV are chosen, and  the colors of lines (black, orange) represent the different values of $\cos\theta$ (0.2, 0.8). We set $u=-2 $ GeV$^2$ in the left panel and $u=-6 $ GeV$^2$ in the right panel. The gaps between the solid and dashed lines are not negligible compared with the magnitude of $d \sigma_{\text{I}}$.

\subsection{Interference term  $d \sigma_{\text{I}}$ with the asymptotic model GDA }

\begin{figure}[ht]
\centering
\includegraphics[width=0.85\textwidth]{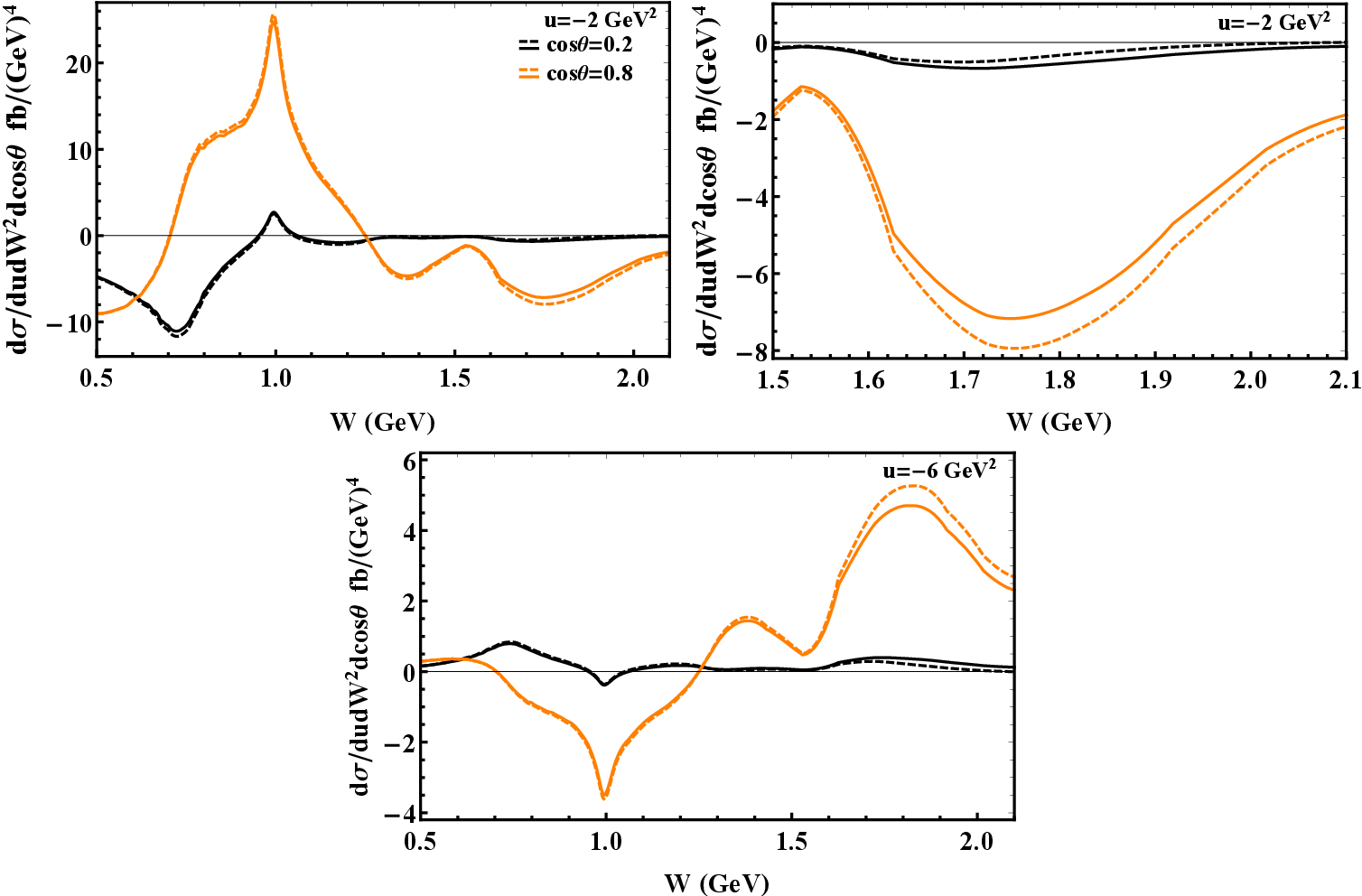}
\caption{ Differential cross section of  the interference term  $d \sigma_{\text{I}}$ at $s=12 $ GeV$^2$,  $u=-2 $ GeV$^2$($-6 $ GeV$^2$).  The  asymptotic $\pi \pi$ GDA is used as the input.}
\label{fig:num5}
\end{figure}

In addition to the extracted $\pi \pi$ GDA, the asymptotic $\pi \pi$ GDA is also used to investigate the kinematical power suppressed corrections in the interference term $d \sigma_{\text{I}}$ of $e^- e^+  \to  \pi^+ \pi^- \gamma$. The asymptotic  GDA is expressed in terms of  the $S$-wave and the $D$-wave terms as explained in Section~\ref{kinematics}, and a model $\pi \pi$ GDA was proposed based on its asymptotic form~\cite{Diehl:1998dk},
\begin{align}
\Phi(z, \cos \theta, \hat{s})=20\, z(1-z)(2z-1) R_{\pi} \left[\frac{-3+\beta_0^2}{2}\, e^{i \delta_0} +
\beta_0^2 e^{i \delta_2}  P_2(\cos \theta)\right],
\label{eqn:gda-asy}
\end{align}
where $\delta_0$  and  $\delta_2$~\cite{Bydzovsky:2016vdx, Bydzovsky:2014cda, Surovtsev:2010cjf} are the  $\pi \pi$ elastic scattering phase shifts of S wave and  D wave, respectively.
The parameter $ R_{\pi}=0.5$ is chosen, which quantifies the  momentum fraction carried by quarks in the pion meson. The asymptotic model GDA is quite different from the extracted $\pi \pi$ GDA. It is thus be meaningful to check whether the kinematical power suppressed corrections are significant or not in $d \sigma_{\text{I}}$  by using both models of the GDAs. In Fig.~\ref{fig:num5}, the interference term  $d \sigma_{\text{I}}$ is depicted by the solid and dashed lines, where the latter are twist-2 cross section and the former include the kinematical power suppressed contribution.  $u=-2 $ GeV$^2$ and $u=-6 $ GeV$^2$ are chosen for  the top and bottom panels, respectively.
The colors (black, orange) of lines represent different values of $\cos \theta$ (0.2, 0.8).
Compared with Fig.~\ref{fig:num3}, both  GDAs predict a similar  magnitude of the interference term $d \sigma_{\text{I}}$. From Fig.~\ref{fig:num5}, we  draw the conclusion that the kinematical contributions cannot be neglected if $W>1$ GeV  which is consistent with the case of the extracted $\pi \pi$ GDA.

\begin{figure}[htp]
\centering
\includegraphics[width=0.85\textwidth]{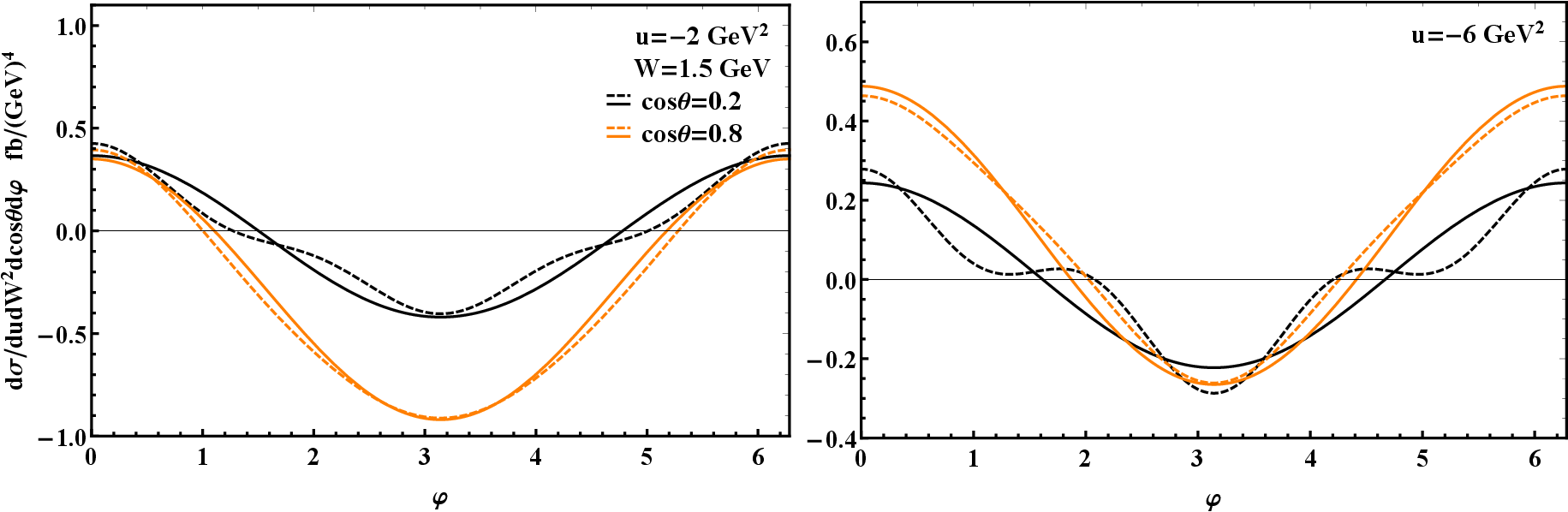}
\caption{Interference term  $d \sigma_{\text{I}}$ is dependent on the azimuthal angle $\varphi$; we  use the same conventions as in Fig.\,\ref{fig:num4a} and the asymptotic model GDA.}
\label{fig:num5a}
\end{figure}

Similarly, we also present the $\varphi$-dependent differential cross section of Eq.\,(\ref{eqn:cro-int}) with the asymptotic model GDA in Fig.\,\ref{fig:num5a}. The kinematics is fixed as $s=12 $ GeV$^2$ and $W=1.5$ GeV, and we choose $u=-2 $ GeV$^2$ ($-6 $ GeV$^2$)for the left (right) panel. The black lines denote $\cos\theta=0.2$, and the orange lines represent $\cos\theta=0.8$. The dashed curves denote the twist-2 cross sections, and the solid ones include the kinematical power suppressed contribution as usual.
Even though the asymptotic model GDA is very different with the extracted $\pi \pi$ GDA especially at $W>1$ GeV (above $KK$ threshold), both  cross sections have similar magnitudes. Moreover,  we also conclude that one cannot neglect the kinematical power suppressed contribution in the $\varphi$-dependent differential cross section.

\section{Summary}
\label{summary}
 GPDs and GDAs are related key physical quantities to study the proton spin puzzle and the  EMT FFs of hadrons, and these research topics  stand at the heart of hadron physics.
 However, there are no experimental facilities where the meson GPDs can be accessed directly, especially the pion meson is one of the Goldstone bosons, thus it is  of prime importance to understand its inner structure. Fortunately, GDAs can be considered as an alternative way to investigate the EMT form factors of mesons, which are accessible in the two-photon reactions, for example $\gamma^{\ast} \gamma \to  M \bar M$ and $\gamma^{\ast} \to M \bar M \gamma$ where the momentum squared of virtual photon is large enough to satisfy the QCD factorization condition. In the former process the meson GDAs are probed by the spacelike photon,
and  the timelike photon  comes from $e^+e^-$ annihilation for the latter. Thus, the universality of GDAs can be checked by two processes \cite{Mueller:2012sma}.

To access the pion GDAs, we investigate the process $e^- e^+ \to \pi \pi \gamma$, and it can be measured at Belle II and BESIII. In case of the $ \pi^0 \pi^0$ production, the cross section is expressed in terms of three helicity amplitudes of $\gamma^*  \to \pi \pi \gamma$ where the GDAs are involved. As indicated by Eq.~(\ref{eqn:epho-cro}), the imaginary phases of the GDAs can not be determined by analyzing its cross section. In the  $ \pi^+ \pi^-$ production channel, $ \pi^+ \pi^-$  can also come from the electromagnetic FF  besides GDAs, which is called as ISR process. Thus, there are three parts in the cross section, among which the interference term between the ISR process and the GDA process is much interesting. The interference term is much larger than the cross section of the GDA process, and it can be obtained if we exchange the momenta of $ \pi^+ \pi^-$, namely $d \sigma(\pi^+\pi^-))-d \sigma(\pi^-\pi^+)=2d \sigma_{\text{I}} $. Therefore, it may be quite easy  measure it. Moreover, the complete information of GDAs can be extracted from the interference term including the imaginary phases.

In this work we provide the complete formula for the interference term in the cross section of the process $e^- e^+  \to \pi^+ \pi^- \gamma$, expressed in terms of three helicity amplitudes and the pion electromagnetic FF. Using the  $\pi \pi$ GDA model extracted from Belle measurements and the pion FF, we also present the numerical estimate of the cross section of $e^- e^+ \to \pi^+ \pi^- \gamma$, and we find that the contribution of ISR process is dominant and the interference term is much larger than the cross section of GDA process, which is consistent with  theoretical expectation.
In order to see the impact of the kinematical power suppressed (kinematical higher-twist) corrections, we calculate the interference term with and without these power suppressed corrections included, using two types of GDAs, which are the extracted $\pi \pi$ GDA and asymptotic model GDA. In case of the measurements of $e^-e^+ \to \pi^+\pi^-\gamma$ at BESIII, our estimates with both GDAs indicate that the kinematical power suppressed corrections contribute significantly to the interference term when $W > 1 $ GeV, and one needs to include such corrections in the extraction of GDAs. If $e^-e^+ \to \pi^+\pi^-\gamma$ is measured at Belle II, the interference term is much smaller than in the case of BESIII. Moreover, the kinematical power suppressed contribution only accounts for $1-2\%$ of the interference term, which also agrees with the rough estimate that
the kinematical contribution is suppressed by the factor of $\hat{s}/s$. Thus, one can adopt the twist-2 formula for the interference term to extract GDAs at Belle II.

As the study of hadron EMT FFs becomes one of the most popular topics in hadron physics, the measurements of $e^- e^+ \to M \bar M \gamma$ and $e^- \gamma \to e^- M \bar M$  are widely discussed for BESIII and Belle II experiments. Actually, such measurements are now  in progress at Belle II. Our theoretical work should play an important role in the precise extraction of meson GDAs from the measurements.
The meson GDAs are used to obtain the meson EMT FFs, from the latter many important physical quantities can be gained, such as mass radius,
mass, pressure and shear force distributions of mesons.

\section*{Acknowledgments}
We acknowledge useful discussions with C\'edric Lorc\'e, Wen-Cheng Yan and Ya-Teng Zhang. BP wants to thank Michel Davier for stimulating discussions on the extraction of GDAs from Babar results almost 10 years ago. Qin-Tao Song was supported by the National Natural Science Foundation
of China under Grant Number 12005191.



\end{document}